\documentclass{appolb}
\usepackage{graphicx}

\begin{document}
\vspace{-2cm}
\title{Recent results from NA61/SHINE and comparison to NA49
\thanks{Presented at the 42 International Symposium on Multiparticle Dynamics (ISMD 2012), Kielce, Poland}%
}
\author{Maja Ma\'{c}kowiak-Paw{\l}owska  for the NA61/SHINE and NA49 Collaborations
\address{Warsaw University of Technology, Faculty of Physics, ul. Koszykowa 75, PL-00-662 Warszawa}
\address{Goethe-Universit\"at Frankfurt am Main, Institut f\"{u}r Kernphysik
Max-von-Laue-Stra\ss e~1
D-60438 Frankfurt am Main}
}
\maketitle
\vspace{-1cm}
\begin{abstract}
The NA61/SHINE experiment aims to discover the critical point of
strongly interacting matter and study properties of the onset of deconfinement. It also performs precise hadron production measurements for the
neutrino and cosmic rays experiments. These goals are being achieved by
measurements of hadron production properties in nucleus-nucleus, proton-proton and proton$/$pion-nucleus interactions as a function of collision energy and
size of the colliding nuclei. 
This contribution presents preliminary results from the NA61 ion program on single-particle spectra and identified particle multiplicity fluctuations in p+p interactions at the CERN SPS. Comparisons with results from p+p and Pb+Pb collisions obtained by the NA49 experiment are shown.
\end{abstract}
\vspace{-0.4cm}
\vspace{-0.4cm} 
\section{Introduction}
The NA61/SHINE experiment and its predecessor NA49 at the CERN SPS study an important region of the phase diagram of strongly interacting matter. The Statistical Model of the Early Stage (SMES) of nucleus-nucleus collisions \cite{SMES} predicted the energy threshold for deconfinement at low SPS energies. Several structures in the excitation functions were expected within the SMES: a kink in the pion yield per participant nucleon (change of slope due to increased entropy production as a consequence of the activation of partonic degrees of freedom), a sharp peak (horn) in the strangeness to entropy ratio, and a step in the inverse slope parameter of transverse mass spectra (constant temperature and pressure in a mixed phase).
These signatures were observed in central Pb+Pb collisions by NA49 around $\sqrt{s_{NN}} = 7.6$~GeV~\cite{NA49}. Fluctuation analysis may provide additional evidence of the onset of deconfinement.\newline 
Moreover, theoretical considerations suggest a critical point (CP) of strongly interacting matter which may be observable
in the SPS energy range according to most lattice QCD calculations~\cite{FodorKatz}. Fluctuations and correlations are basic tools to study this phenomenon. Enlarged fluctuations are expected close to the critical point. In nucleus-nucleus collisions a maximum of fluctuations is expected when freeze-out happens near the CP.

\vspace{-0.4cm}
\section{The NA61/SHINE experiment}

\begin{figure}[htb]
\begin{center}
\vspace{-0.6cm}
\includegraphics[width=2.9in]{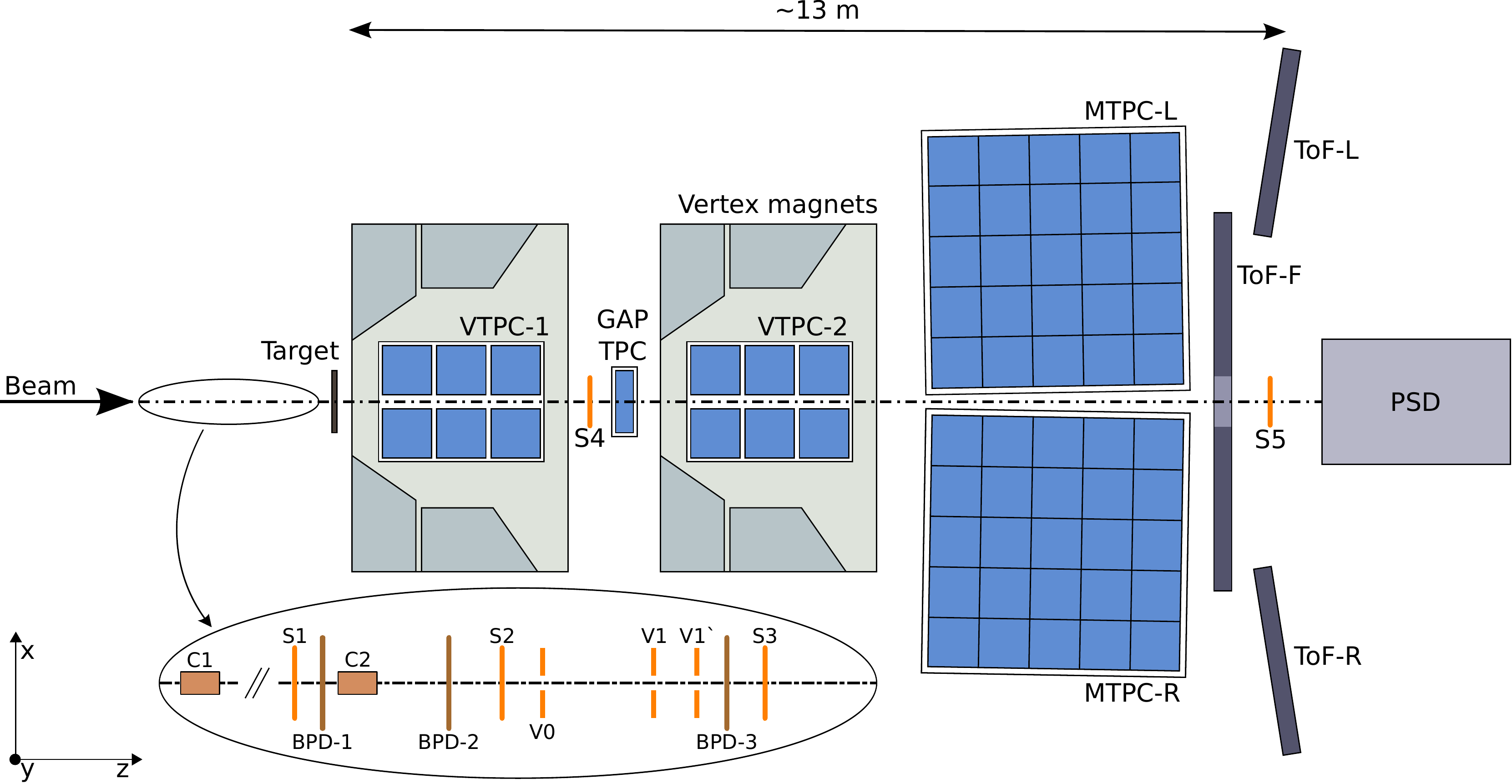}
\includegraphics[width=2in]{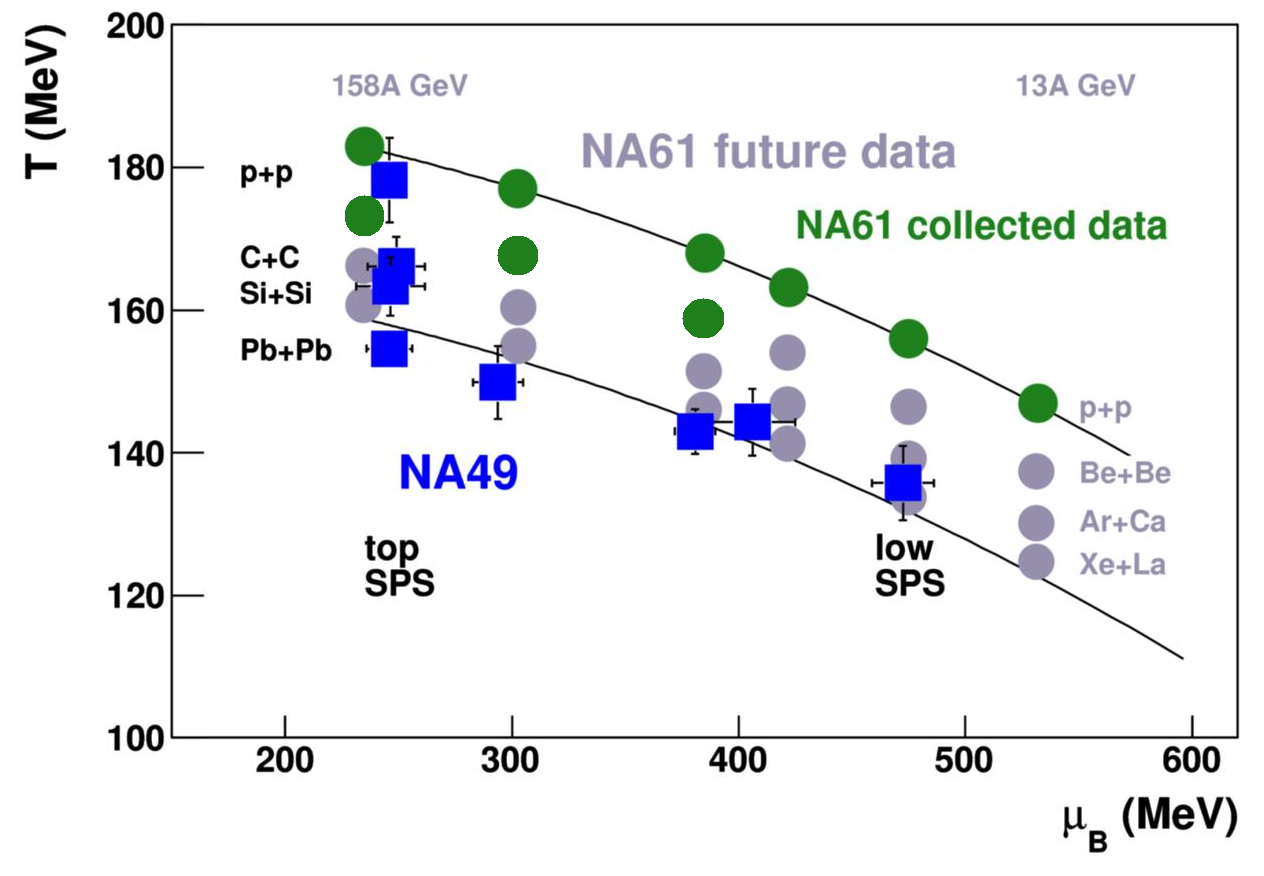}
\end{center}
\vspace{-0.6cm}
\caption{The NA61 setup (left) and the energy - system size scan of NA61 in progress (right). Blue rectangles indicate data collected by NA49, circles indicate planned NA61 data, green circles are data points already recorded by NA61.}
\label{na61setup}
\end{figure}

NA61/SHINE is a fixed target experiment located in the North Area of CERN. {\it SHINE} stands for {\bf S}PS {\bf H}eavy {\bf I}on and {\bf N}eutrino {\bf E}xperiment. It was approved in 2007. NA61 greatly profits from the long development of the CERN particle sources and the accelerator chain as well as the H2 beam line of the CERN North Area. The latter has recently been modified to also serve as a fragment separator as needed to produce the Be beam for NA61. NA61/SHINE inherited the basic components from its predecessor, the NA49 experiment, and added important upgrades. The NA61 setup is presented on Fig.~\ref{na61setup}~(left).\newline

The detector consists of five Time-Projection Chambers. Three of them (VTPC-1, VTPC-2~and~GAP-TPC) are located in a magnetic field. Two others (MTPC-L and MTPC-R) are located downstream of the VTPCs. The TPCs are supplemented by three Time-of-Flight walls located behind the MTPCs. This setup allows for good particle identification and momentum resolution.\newline
The NA61$/$SHINE goal is to measure hadron production in p+p, p+A, h+A and A+A interactions at the SPS energies, in order to:
\begin{itemize}
	\item obtain precise data on hadron production (spectra) in p+C interactions at 31 GeV$/$c for the T2K experiment for precisely computing the initial neutrino flux from the T2K target at J-PARC,
	\item perform reference measurements of $\pi$+C and p+p interactions for cosmic-ray physics (Pierre-Auger and KASCADE experiments) for improving air shower simulations,
	\item search for the critical point of strongly interacting matter,
	\item study the properties of the onset of deconfinement,
	\item study of high $p_{T}$ particle production (energy dependence of $R_{AA}$).
\end{itemize}
In order to study properties of the onset of deconfinement and search for the critical point NA61 performs for the first time in history a two dimensional scan in system size and energy. Already gathered systems and energies as well as planned measurements are presented in 
Fig.~\ref{na61setup} (right) in the phase diagram of temperature and baryo-chemical potential.

\section{NA61/SHINE measurements for the onset of deconfinement and critical point}
The following results were obtained from p+p data collected in 2009 by NA61/SHINE at $\sqrt{s_{NN}} = (6.2\footnote{spectra only})7.6 - 17.3$ GeV. Event and track cuts were chosen to select only inelastic interactions and charged particles produced directly in the interactions or via strong and electromagnetic decays.

\subsection{NA61 results on single-particle spectra}
\begin{figure}[htb]
\begin{center}
\vspace{-0.6cm}
\includegraphics[width=2.3in]{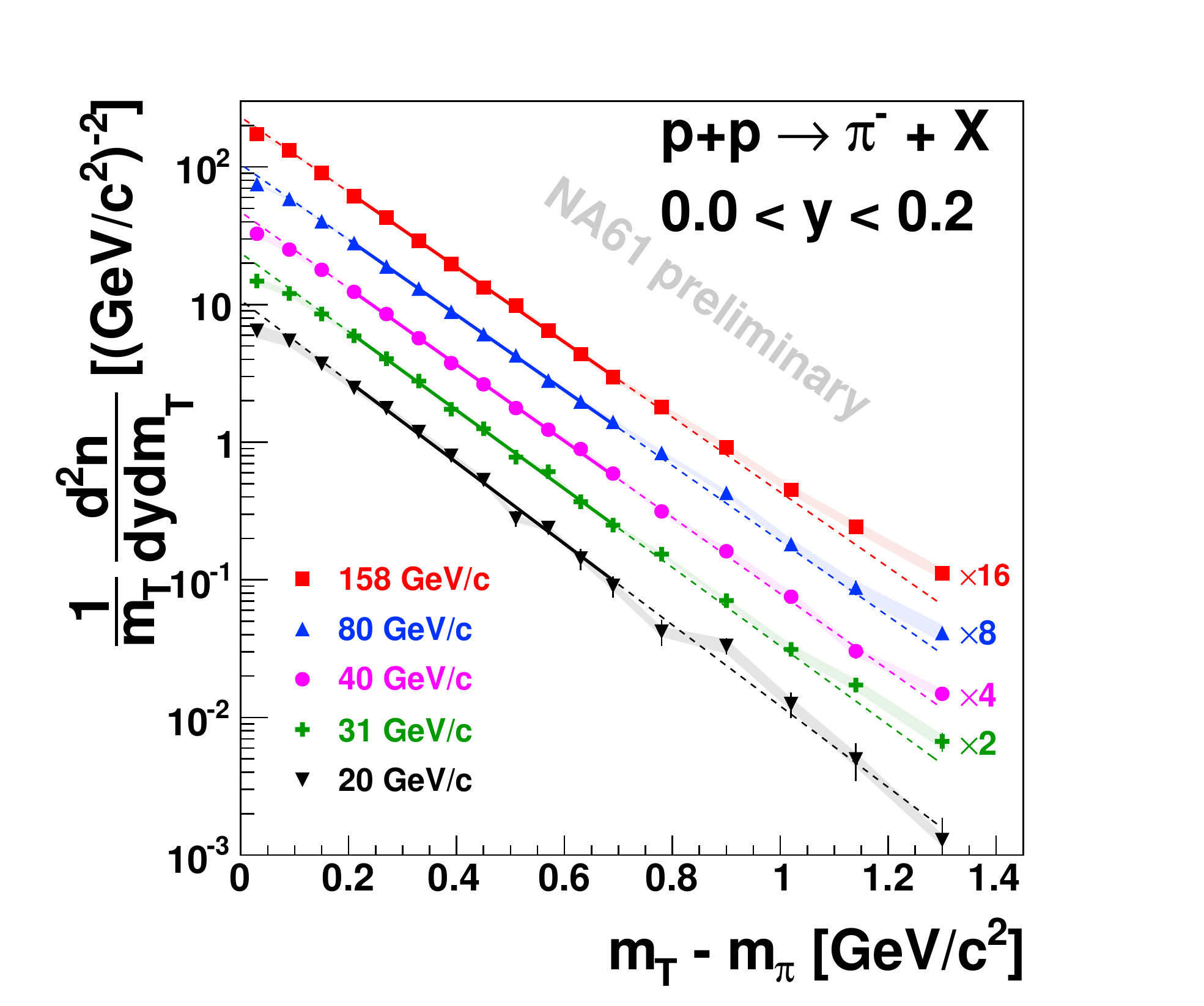}
\includegraphics[width=2.3in]{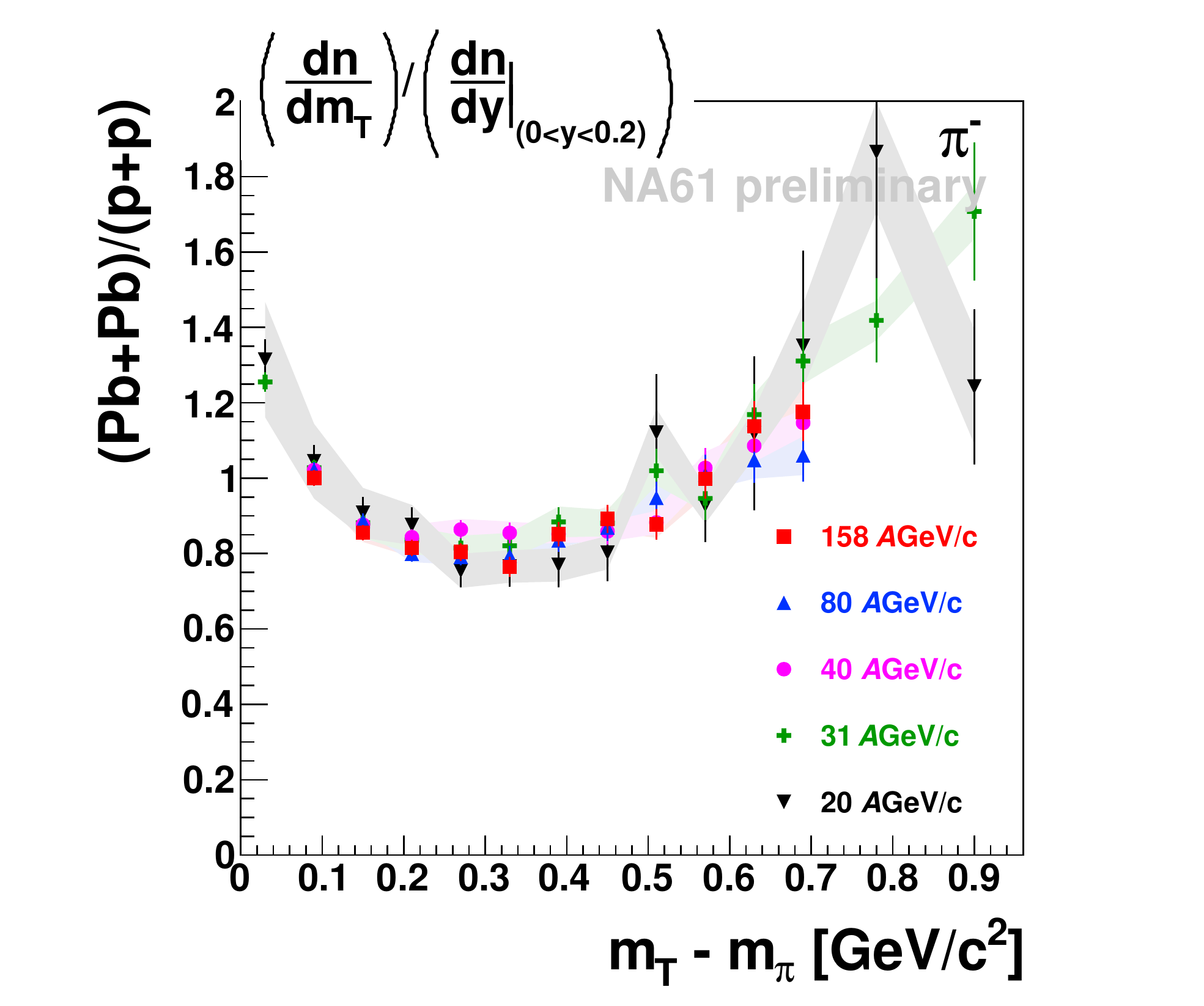}
\end{center}
\vspace{-0.6cm}
\caption{Transverse mass spectra of $\pi^{-}$ mesons in inelastic p+p interactions at $6.2-17.3$~GeV (left) compared to central Pb+Pb data of NA49 (right). Statistical errors are indicated by vertical bars and systematic uncertainties by colored bands.}
\label{mt}
\end{figure}

\begin{figure}[htb]
\begin{center}
\vspace{-0.2cm}
\includegraphics[width=2.2in]{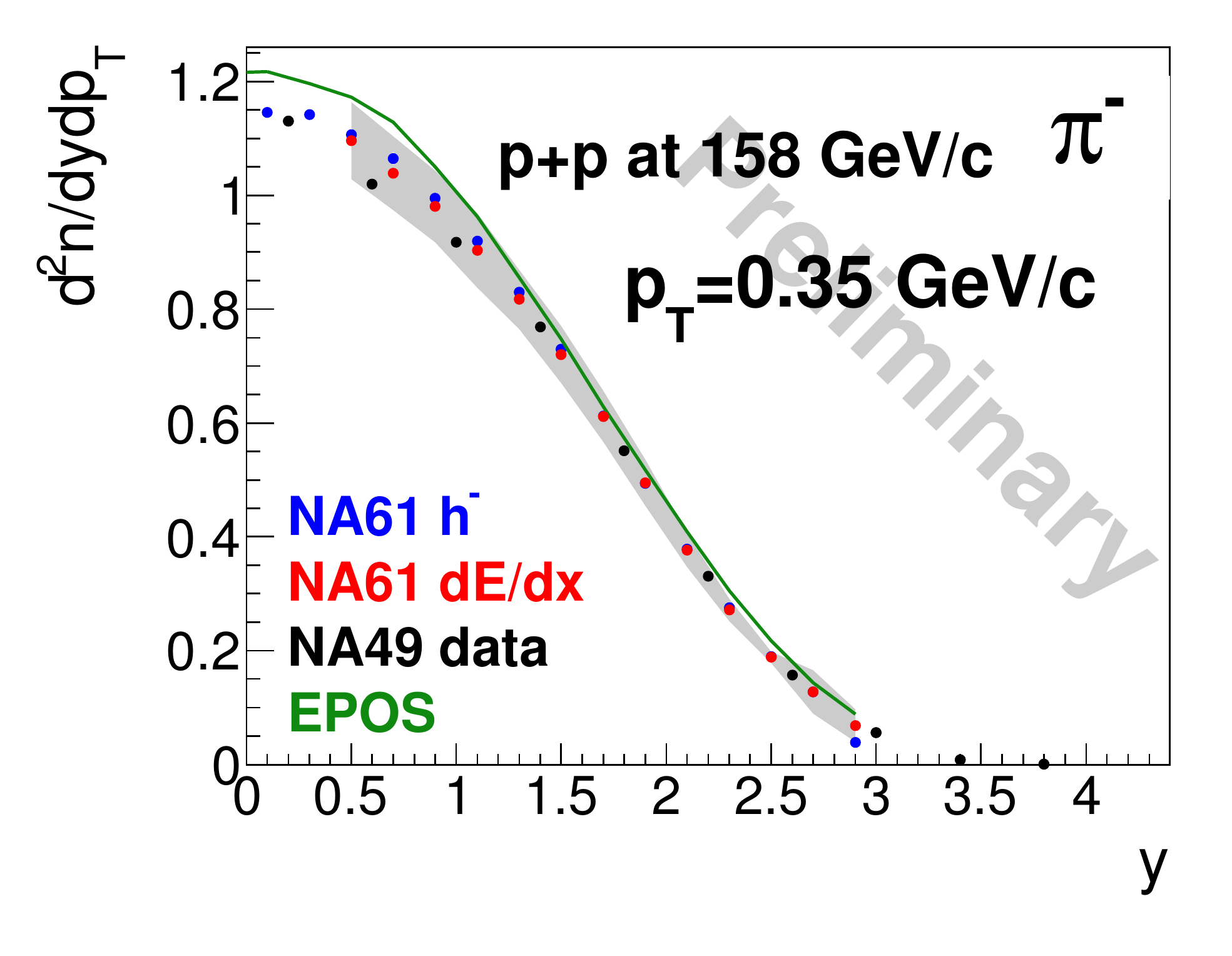}
\includegraphics[width=2.2in]{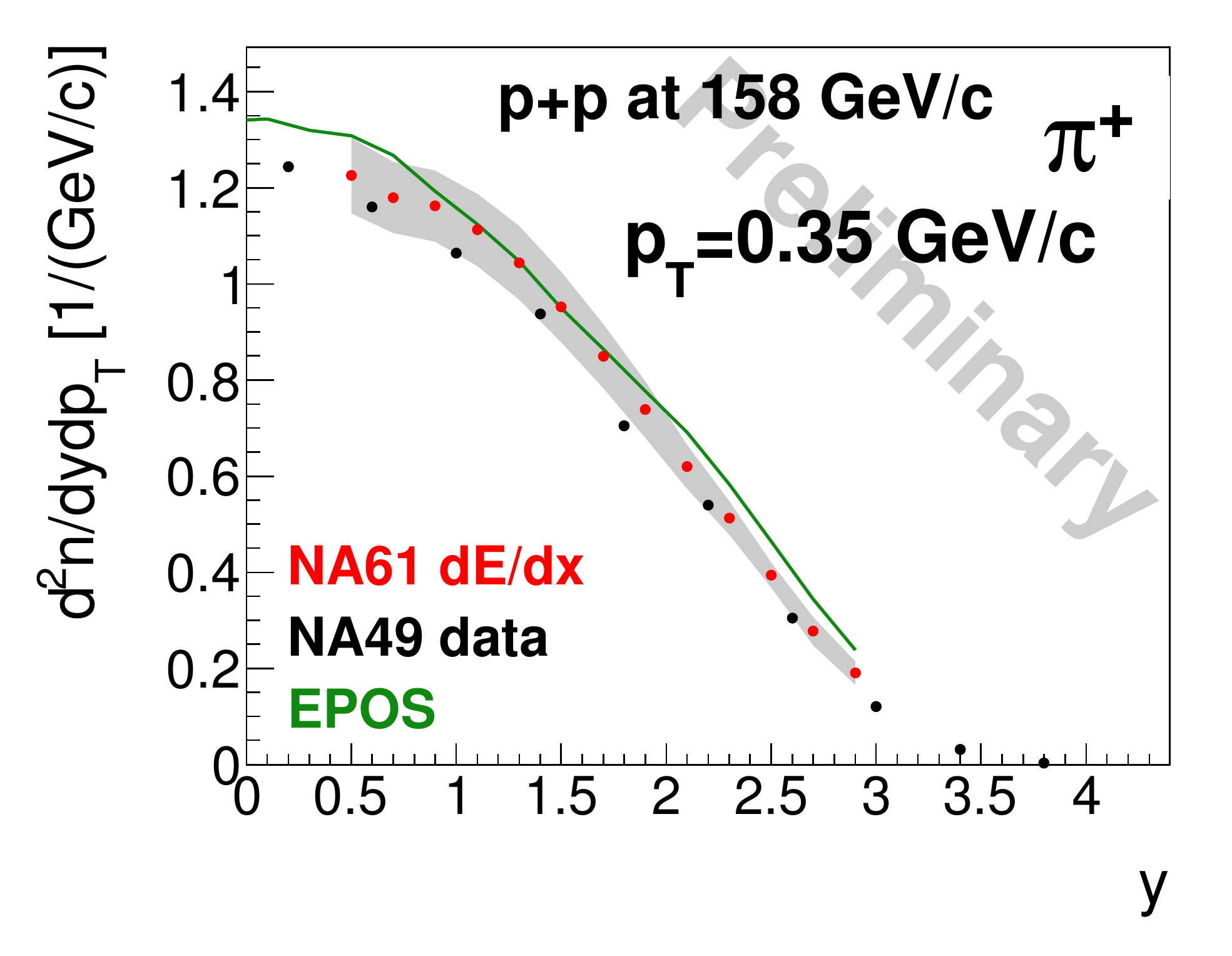}
\includegraphics[width=2.2in]{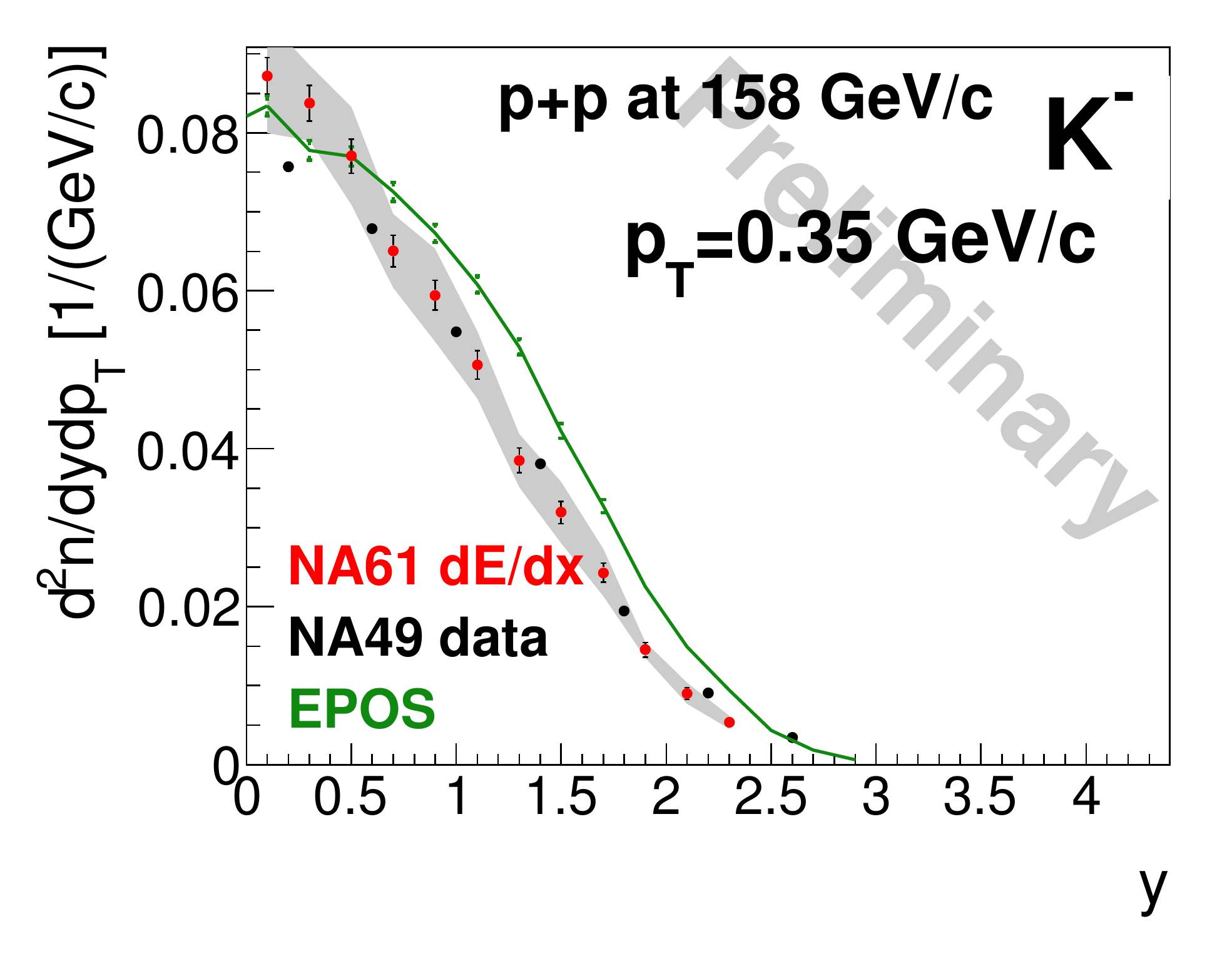}
\includegraphics[width=2.2in]{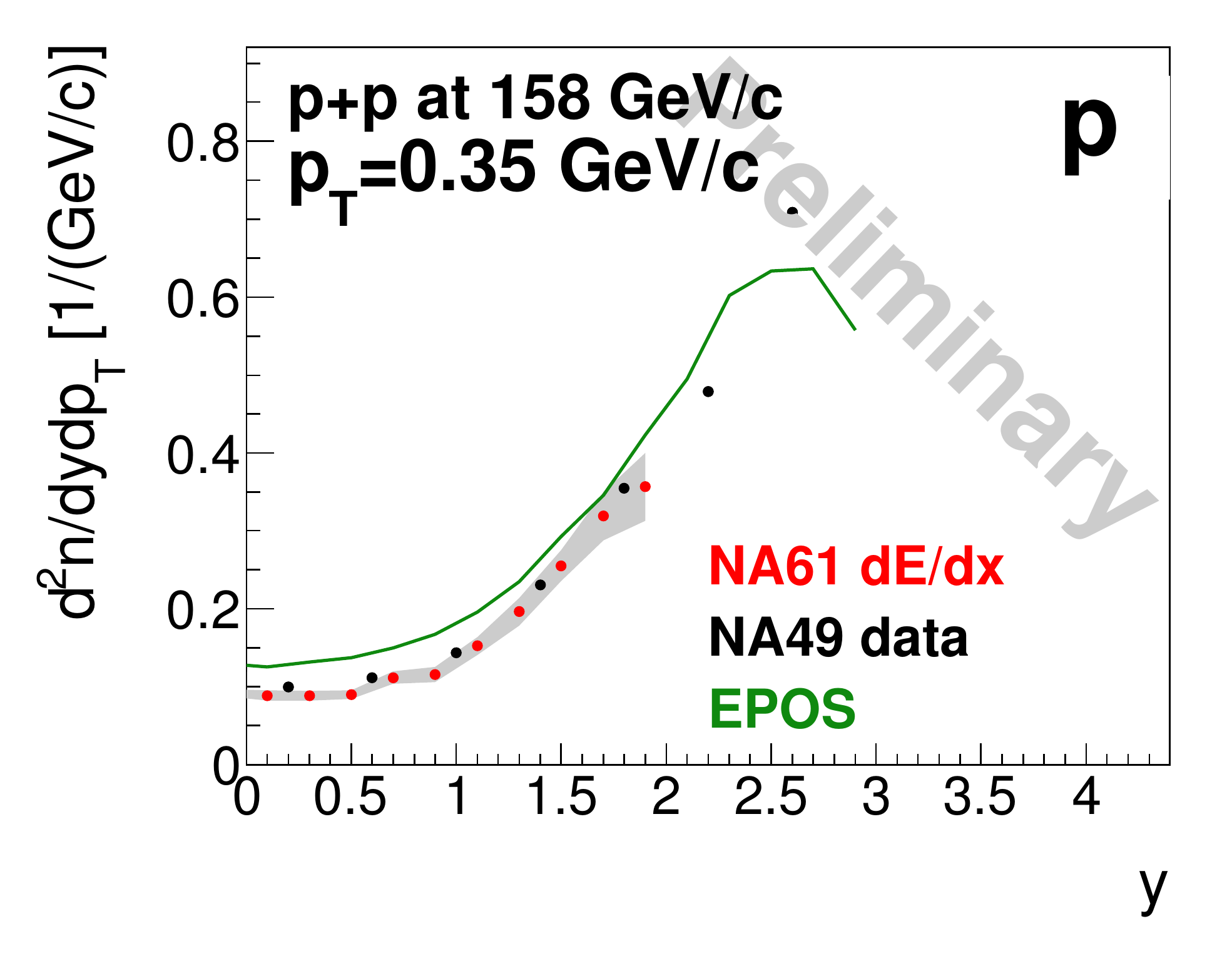}
\end{center}
\vspace{-0.6cm}
\caption{Rapidity spectra of $\pi^{-}$, $\pi^{+}$, $K^{-}$ mesons and protons in inelastic p+p interactions at 17.3 GeV. The NA61 results are compared with the corresponding NA49 data and predictions of the EPOS model.}
\label{Szymon}
\end{figure}

Results on single-particle spectra for $\pi^{-}$ mesons are obtained using two methods. First, by the $h^{-}$ method which is based on the fact that the majority of negatively charged particles are $\pi^{-}$ mesons. The contribution of other particles was estimated and subtracted using the VENUS and EPOS models. Second, by the $dE/dx$ method which uses information on particle energy loss in the TPC gas in the relativistic rise region to identify particles. Results from both methods are corrected for feed-down from weak decays and detector effects using simulations. Out-of-target interactions are subtracted using events recorded with the empty liquid hydrogen target.\newline 
Transverse mass spectra of $\pi^{-}$ in p+p interactions at $6.2-17.3$~GeV are presented in Fig.~\ref{mt} (left). For all energies $m_{T}$ spectra are approximately exponential. The ratio of NA49 spectra for central Pb+Pb collisions \cite{PbPbna491,PbPbna492} to those of NA61 for p+p interactions, both normalized to unity, is shown in Fig.~\ref{mt} (right). The concave shape of the ratio seems to be energy independent.\newline
\begin{figure}[htb]
\begin{center}
\vspace{-0.6cm}
\includegraphics[width=2.4in]{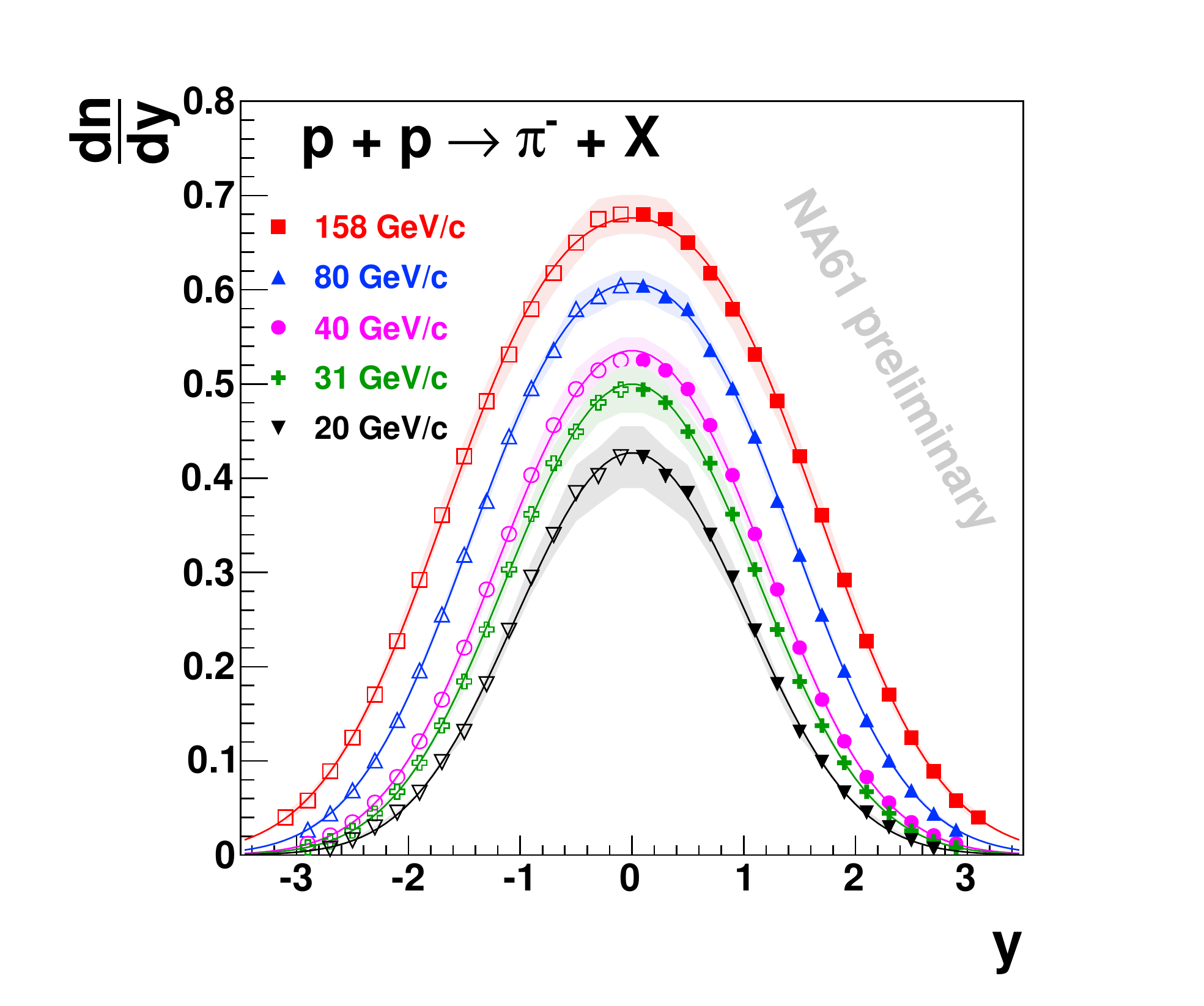}
\includegraphics[width=2.4in]{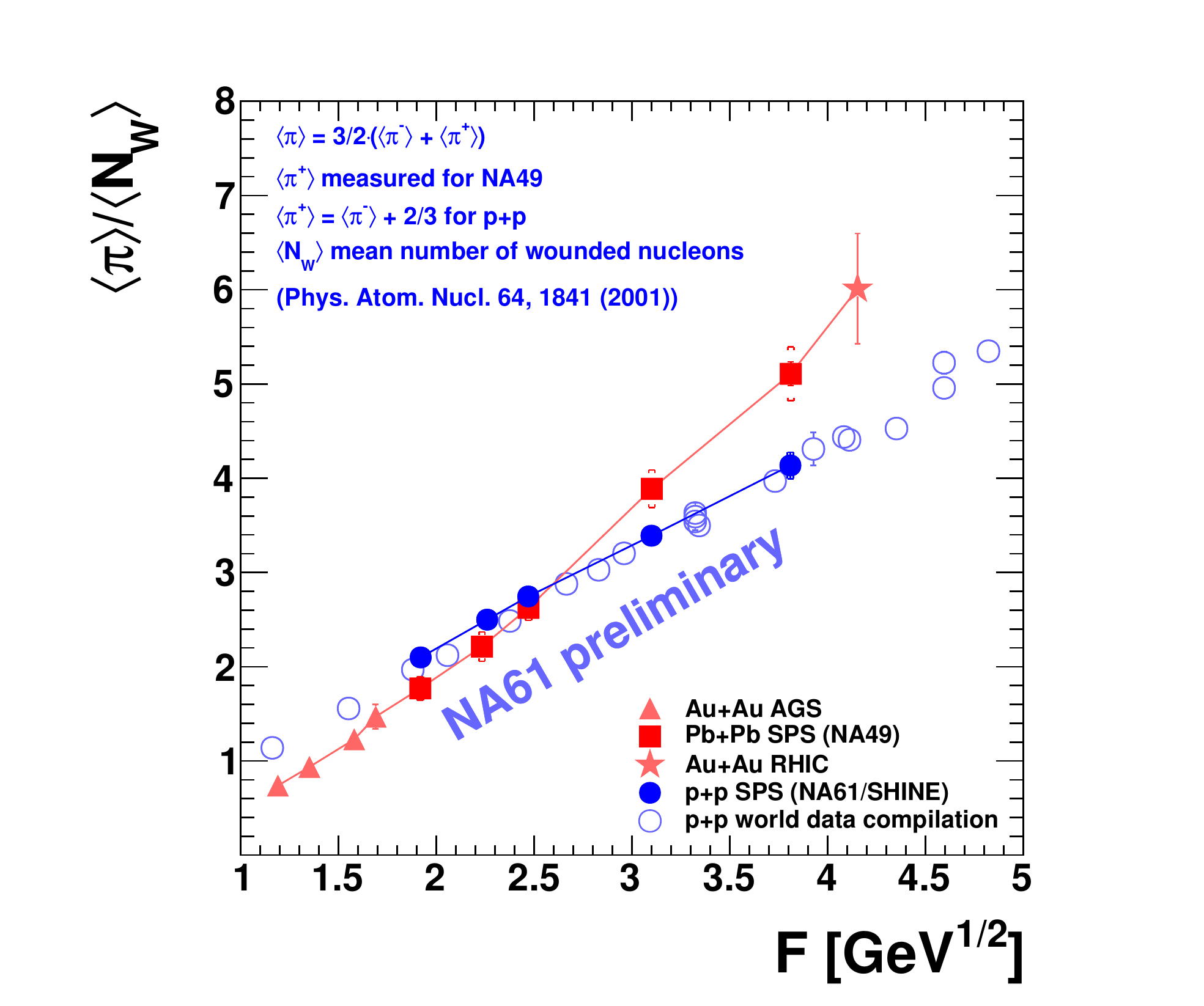}
\end{center}
\vspace{-0.6cm}
\caption{Rapidity spectra of $\pi^{-}$ mesons in inelastic p+p interactions at $6.2-17.3$ GeV (left) and mean pion multiplicity per wounded nucleon as a function of the Fermi energy measure $F=\frac{(\sqrt{s_{NN}}-2\cdot m_{N})^{3/4}}{(\sqrt{s_{NN}})^{1/4}}\approx \sqrt{\sqrt{s_{NN}}}$ (right).}
\label{kink}
\end{figure} 
Rapidity spectra of $\pi^{-}$ mesons in inelastic p+p interactions at $6.2-17.3$~GeV are presented in Fig.~\ref{kink} (left). An excellent fit at all energies is obtained by the sum of two symmetrically displaced Gaussian functions
from which the mean $\pi^{-}$ multiplicity was derived.\newline
The mean yield of pions ($\pi^{+}+\pi^{-}+\pi^{0}$) per wounded nucleon (Fig.~\ref{kink} (right)) was
calculated using phenomenological isospin factors \cite{piony}. NA61/SHINE measurements are in agreement with the world data. The mean multiplicity in central Pb+Pb rises faster than in p+p collisions with a crossover at about $\sqrt{s_{NN}}=10$~GeV. The precision of the data allows for a detailed study of the onset of deconfinement.\newline
Rapidity spectra of $\pi^{-}$, $\pi^{+}$, $K^{-}$ mesons and protons produced in inelastic p+p interactions at $17.3$~GeV are presented in Fig.~\ref{Szymon}. The spectra are compared with the corresponding NA49 results \cite{na491,na492,na493}\footnote{NA49 points are obtained by interpolation from $x_{F}-p_{T}$ spectra} as well as with the EPOS model \cite{EPOS}. The experimental data agree for all particle types, between both analysis methods and NA61 and NA49. The EPOS model shows significant deviations from the data for $K^{-}$ mesons and protons.

\subsection{NA61 results on chemical fluctuations}
Particle identification for the chemical fluctuation analysis is based on energy loss measurements in the relativistic rise region. It does not allow unique identification of hadrons. The mean hadron multiplicities are obtained by fitting a sum of response functions for each particle type to the measured inclusive dE$/$dx distributions in narrow phase space bins. In order to analyze fluctuations of identified hadrons in p+p interactions of NA61/SHINE, as well as Pb+Pb collisions of NA49 a new approach, the so-called identity method \cite{identity1,identity2,identity3}, was used. It allows to obtain second and third moments (pure and mixed) of identified particle multiplicity distributions corrected for the effect of imprecise event-by-event particle identification.  \newline
\begin{figure}[htb]
\begin{center}
\includegraphics[width=2.1in]{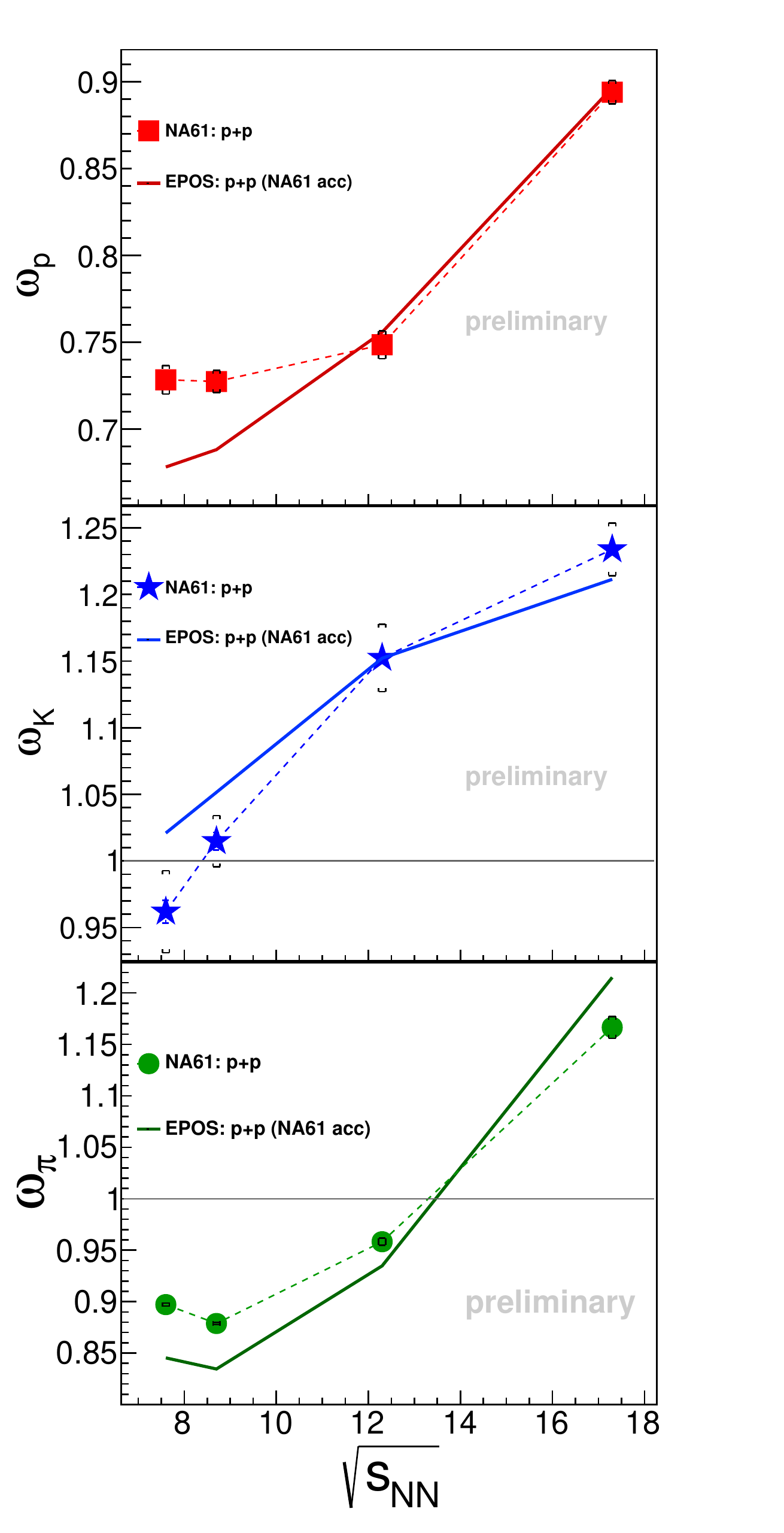}
\includegraphics[width=2.2in]{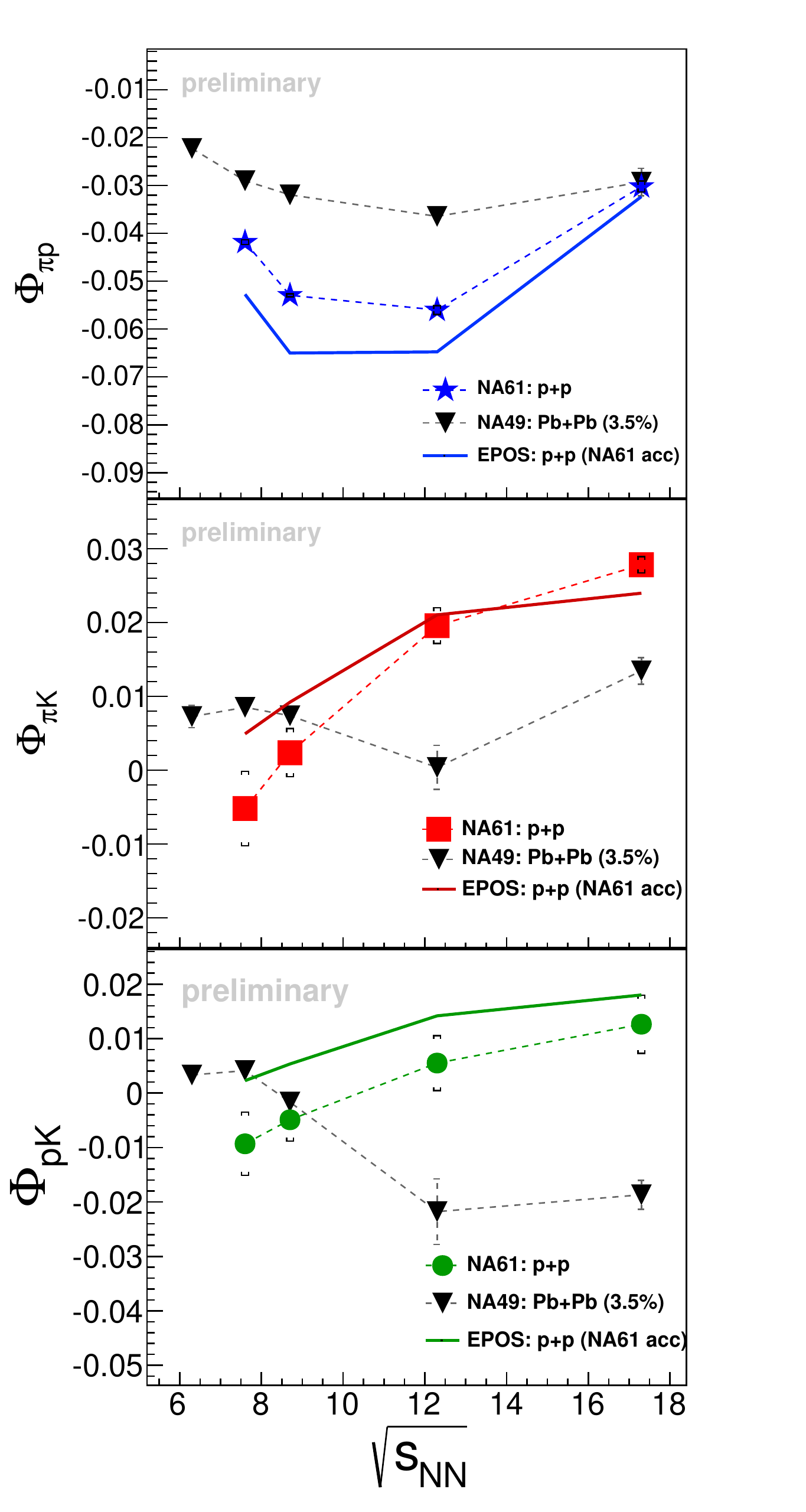}
\end{center}
\vspace{-0.6cm}
\caption{Collision energy dependence of the scaled variance of identified hadrons multiplicity distributions in inelastic p+p interactions (left). The $\Phi_{ij}$ fluctuation measure as a function of collision energy for p+p (NA61) and central Pb+Pb (NA49) collisions (right). The p+p data are compared with the EPOS model predictions.}
\label{chemfluct}
\end{figure}
Two quantities, scaled variance $\omega_{i}$ of the multiplicity distribution and the two-particle measure $\Phi_{ij}$, were chosen for the chemical fluctuation analysis. The scaled variance is defined as:
\begin{equation}
	\omega_{i} = \frac{<N_{i}^{2}>-<N_{i}>^{2}}{<N_{i}>} ,
\end{equation}
where $<N_{i}>$ and $<N_{i}>^{2}$ are the mean multiplicity and second moment of the multiplicity distribution of particles of type {\it i}, respectively.\newline
The scaled variance is an intensive measure \cite{SIQ} i.e. it is independent of the number of wounded nucleons in the Wounded Nucleon Model or volume in the Grand Canonical Ensemble but it depends on their fluctuations. The last feature makes it difficult to compare p+p interactions with nucleus-nucleus collisions. For the Poisson distribution $\omega=1$.\newline
Figure \ref{chemfluct} (right) shows the scaled variance for $\pi=\pi^{+}+\pi^{-}$, $K=K^{+}+K^{-}$ and $p=p+\bar{p}$ as a function of collision energy. Results refer to the NA61 acceptance for chemical fluctuation analysis \cite{akceptacja}.\newline
The scaled variance for all studied hadron types increases with increasing collision energy. The trend as well as the magnitude of the effect is well reproduced by the EPOS model. As pions are the most abundant particles their fluctuations should be dominated by the KNO scaling \cite{KNO}. It yields a linear increase of the scaled variance with mean multiplicity of charged particles. In NA61 this trend is modified by the limited detector acceptance. \newline
Values of $\omega_{K}$ for all energies are close to or above one. This is most probably caused by strangeness conservation which leads to a correlation between the production of $K^{+}$ and $K^{-}$ mesons. Again the effect is weakened by the limited acceptance. To the contrary baryon number conservation seems to suppress proton fluctuations. This is because the proton multiplicity is mostly given by the two initial protons as production of proton-antiproton pairs is strongly suppressed by their large masses. This suppression effect is again weakened by the limited acceptance and it is expected to decrease with increasing energy. \newline
In order to compare results for p+p and central Pb+Pb collisions, the strongly intensive measure $\Phi_{ij}$ \cite{SIQ,Phi} defined for two hadron types, i and j, was chosen. It is defined as:
\begin{equation}
	\Phi_{ij}=\frac{\sqrt{<N_{i}><N_{j}>}}{<N_{i}+N_{j}>}\cdot[\sqrt{\Sigma^{ij}}-1] ,
\end{equation}
where  $\Sigma^{ij}=[<N_{i}>\cdot~\omega_{j}~+~<N_{j}>~\cdot~\omega_{i}~-2\cdot(<N_{ij}>-<N_{i}><N_{j}>)] / <N_{i}+N_{j}>$. As a strongly intensive measure, $\Phi_{ij}$ is not only independent of number of wounded nucleons or volume but also of their fluctuations.\newline
Figure \ref{chemfluct} shows the energy dependence of $\Phi_{ij}$ for combinations of two hadron types: $\pi p$, $\pi K$ and $pK$. When no 
inter-particle correlations are present $\Phi_{ij}=0$.\newline
For $\pi K$ and $p K$ $\Phi_{ij}$ increases with increasing energy. For $\Phi_{\pi p}$ there is a minimum between $7.3$ and $8.7$ GeV. A similar but weaker effect is visible in Pb+Pb interactions. It also appears in the EPOS model. The increase of $\Phi_{\pi K}$ for p+p interactions is not observed in Pb+Pb collisions. $\Phi_{pK}$ shows a clear difference between results for p+p, which increase with increasing energy, and Pb+Pb which decrease with increasing energy. Both dependencies cross zero at the same energy $\sqrt{s_{NN}}\approx8.7$ GeV.

\vspace{-0.3cm}
\section{Conclusions}
The NA61/SHINE experiment successfully started the planned 2D energy-system size scan with p+p interactions. The results are needed as reference for the onset of deconfinement study and the search for the critical point.\newline
The $m_{T}$ spectra of $\pi^{-}$ mesons are approximately exponential. The comparison to Pb+Pb interactions shows noticeable differences for the lowest and highest $m_{T}$ values which seem to be energy independent. A sum of two Gaussian functions is required to fit the rapidity spectra of $\pi^{-}$.\newline
The energy dependence of the mean pion multiplicity per wounded nucleon from NA61/SHINE agrees well with the previously established trend.\newline
Fluctuations of identified $\pi$, $K$ and protons in inelastic p+p interactions are well described by the EPOS model. Conservation laws seem to play an important role in fluctuations of hadron multiplicities. Comparison with data on central Pb+Pb collisions measured by NA49 shows large differences between p+p and Pb+Pb in terms of fluctuations which are most pronounced for $p K$ fluctuations.
\vspace{-0.2cm}
\begin{center}
Acknowledgments
\end{center}
This work was partially supported by the National Center of Science under grant
DEC-2011/03/B/ST2/02617 and the German Research Foundation (DFG grant GA 1480/2.1).


\vspace{-0.6cm}

\begin{thebibliography}{9}

\bibitem{SMES}
  M. Ga\'{z}dzicki and M. I. Gorenstein, Acta\ Phys.\ Polon.\  B {\bf 30}, 2705 (1999)
\bibitem{NA49}
 NA49 Coll., Phys.\ Rev.\ C {\bf 77}, 024903 (2008)
\bibitem{FodorKatz}
  Z. Fodor and S. D. Katz, JHEP {\bf 0404}, 050 (2004)
\bibitem{PbPbna491}
  NA49 Coll., Phys.\ Rev.\ C {\bf 66}, 054902 (2002)
\bibitem{PbPbna492}
  NA49 Coll., Phys.\ Rev.\ C {\bf 77}, 024903 (2008)
\bibitem{piony}
 A.I. Golokhvastov, Phys.\ Atom.\ Nucl.\ {\bf 64}, 1841 (2001)
\bibitem{na491}
	NA49 Coll., Eur. Phys. J. C {\bf 68}, 1-73 (2010) 
\bibitem{na492}
	NA49 Coll., Eur. Phys. J. C {\bf 65}, 9-63 (2010) 
\bibitem{na493}
	NA49 Coll., Eur. Phys. J. C {\bf 45} (2006)

\bibitem{EPOS}
	K. Werner et al., Phys.\ Rev.\ C {\bf 74}, 044902 (2006)

\bibitem{identity1} 
  M.~Gazdzicki, K.~Grebieszkow, M.~Mackowiak and S.~Mrowczynski,
  Phys.\ Rev.\ C {\bf 83}, 054907 (2011)

\bibitem{identity2} 
  M.~I.~Gorenstein,
  Phys.\ Rev.\ C {\bf 84}, 024902 (2011)


\bibitem{identity3} 
  A.~Rustamov and M.~I.~Gorenstein,
  Phys.\ Rev.\ C {\bf 86}, 044906 (2012)

\bibitem{SIQ} 
  M.~I.~Gorenstein and M.~Gazdzicki,
  Phys.\ Rev.\ C {\bf 84}, 014904 (2011)
 
\bibitem{akceptacja}
https://edms.cern.ch/document/1237791/1

\bibitem{KNO}
Z. Koba, H. B. Nielsen and P. Olesen, Nucl.\ Phys.\ B {\bf 40}, 317 (1972) 

\bibitem{Phi} 
  M.~Gazdzicki,
  Eur.\ Phys.\ J.\ C {\bf 8}, 131 (1999)


\end{thebibliography}
\end{document}